\documentstyle[mprocl,epsfig]{article}

\def\be{\begin{equation}}
\def\ee{\end{equation}}
\def\bea{\begin{eqnarray}}
\def\eea{\end{eqnarray}}

\begin{document}

\title{Non-local current correlations in ferromagnet/superconductor
nanojunctions}

\author{C. J. Lambert\thanks{e-mail:c.lambert@lancaster.ac.uk}}
\address{Department of Physics, Lancaster University,
Lancaster, LA1 4YB, UK}
\author{J. Koltai, J. Cserti\thanks{e-mail: cserti@galahad.elte.hu},}
\address{Department of Physics of Complex Systems,
E{\"o}tv{\"o}s University
\\ H-1117 Budapest, Hungary}
\date{\today}
\maketitle\abstracts{
 When two fully polarized
ferromagnetic (F) wires with opposite polarizations make contact
with a spin-singlet superconductor, a potential-induced current in
wire 1 induces a non-local current of equal magnitude and sign in
wire 2. The magnitude of this current has been studied in the
tunneling limit and found to decay exponentially with the
distance between the contact. In this paper we propose a new
structure in which this novel non-local effect is increased by
orders of magnitude. We study the spin-dependent electronic
transport of a diffusive nanojunction and demonstrate that when a
normal diffusive region is placed between the F leads and
superconductor, the non-local initially increases with the
separation between the F leads, achieving a maximum and decays as
a power law with increasing separation.}



\vspace{5mm}

Experimental studies of electronic transport properties of
nanostructures containing both ferromagnets (F) and
superconductors (S)
 \cite{Upd1,Upd2,Soul,Fierz,Gior1,vasco,Pann,Petra,Gior2,jed,Bour}.
reveal novel features, not present in normal-metal/superconductor
(N/S) junctions, due to the suppression of electron-hole
correlations in the ferromagnet. When spin-flip processes are
absent, further effects are predicted, including the suppression
of the conventional giant magnetoresistance ratio in diffusive
magnetic multilayers \cite{gmr} and the appearance of non-local
currents when two fully polarized ferromagnetic (F) wires with
opposite polarizations make contact with a spin-singlet
superconductor \cite{feinberg,feinberg1}. The latter effect has
been highlighted, because of interest in the possibility of
generating entangled pairs of electrons at a superconductor (S)
interface \cite{tangle1,tangle2,tangle3}. A recent study of such a
junction in the tunneling limit predicts that the magnitude of
the non-local current will decrease exponentially with the
distance between the F contacts. It is therefore of interest to
ask how the magnitude of this novel effect can be enhanced.

In this paper we propose a hybrid nanostructure in which the
non-local current is enhanced by orders of magnitude compared
with the geometry of ref.  \cite{feinberg}. The proposed structure
is sketched in figure 1 and comprises two clean F wires, each of
width $M_f$, separated by a distance $M$, in contact with a
diffusive, normal metallic region of area $A=L.(2M_f+M)$, which
in turn makes contact with a spin-singlet superconductor.

In the linear response limit, the current-voltage relation of a
such a hybrid structure connected by non-superconducting wires to
normal reservoirs was first presented in ref \cite{lam1}. If $I_1,
I_2$ are the currents leaving reservoirs 1 and 2, and $v_1, v_2$
their respective voltages, then at zero temperature \cite{lam1}

$$I_1 =[2e^2/h][(N- R_0+R_a)(v_1-v) +
(T^\prime_a-T^\prime_0)(v_2-v)]\eqno{(1a)}$$ and
$$I_2 =[2e^2/h][
(T_a-T_0)(v_1-v)+(N- R^\prime_0+R^\prime_a)(v_2-v)],\eqno{(1b)}$$
where $v$ is the condensate potential.
 In this expression $R_0$ is
the coefficient for an electron from reservoir 1 to be reflected
as an electron back into reservoir 1, while $T_0$ is the
coefficient for an electron from reservoir 1 to be transmitted as
an electron into reservoir 2. $R_a$ is the coefficient for an
electron from reservoir 1 to be Andreev reflected as a hole into
wire 1 and $T_a$ is the coefficient for an electron from
reservoir 1 to be transmitted as a hole into wire 2. Finally
$R^\prime_0,\,\, T^\prime_0,\,\, R^\prime_a,\,\, T^\prime_a$ are
the corresponding coefficients for electrons originating from
reservoir 2. All coefficients are evaluated at the Fermi energy
and are obtained by summing over the elements of sub-matrices of
the quantum mechanical scattering matrix $S$.  As a consequence of
unitarity of $S$, they satisfy $N=
R_0+R_a+T_a+T_0=R^\prime_0+R^\prime_a+T^\prime_a+T^\prime_0$,
where $N$ is the number of open scattering channels in the wires.

As discussed in ref  \cite{lam2}, equations (1) are a convenient
starting point for describing a wide variety of transport
phenomena in hybrid superconducting nanostructures.  For example
if lead 2 is employed as a voltage probe, with $I_2$ set to zero,
then equation (1b) yields
$$(v_2-v)/(v_1-v)=(N- R^\prime_0+R^\prime_a)/(T_0-T_a).\eqno{(2)}$$
 In the
absence of Andreev processes, this ratio is positive. However if
Andreev transmission dominates normal transmission, the ratio can
be negative, leading to novel negative 4-probe conductances in
hybrid superconducting structures \cite{lam4probe}. The appearance
of non-local currents is similarly obtained from equations (1) by
setting $v_2-v=0$, which yields
$$I_1=[2e^2/h][N-R_0+R_a](v_1-v)=[2e^2/h][2R_a+T_0+T_a](v_1-v)\eqno{(3)}$$
and
$$I_2=[2e^2/h][T_a-T_0](v_1-v).\eqno{(4)}$$
When all Andreev processes are absent, this yields the expected
result $I_2=-I_1=-(2e^2/h)T_0(v_1-v)$, whereas when Andreev
reflection and normal transmission are completely suppressed,
$I_2=+I_1=(2e^2/h)T_a(v_1-v)$. The latter occurs when the F wires
are completely spin polarized and the polarization in wire 1 is
the opposite of that in wire 2. More generally $I_1$ and $I_2$
will have opposite signs whenever Andreev transmission dominates
normal transmission. As noted in ref \cite{lam4probe}, this can occur
even if the current carrying wires are not ferromagnetic.

In the absence of a normal region in front of the superconductor
(ie $L=0$) and in the presence of tunnel junctions at the F-S
interfaces, the analysis of ref \cite{feinberg} predicts that both
$I_1$ and $I_2$ decay as $\exp-(2L/\pi\xi)$, where $\xi$ is the
superconducting coherence length. To analyze this structure for
finite $L$ we solve the Bogoliubov - de Gennes equation on a
square tight binding lattice and compute the scattering matrix
using an exact recursive Green's function technique.

We start by defining a tight-binding lattice of sites with the
geometry of figure 1. Each site is labeled by a lattice vector
$\vec l$ and possesses particle (hole) degrees of freedom
$\psi^\sigma(\vec l)$ $(\phi^\sigma(\vec l))$, with spin
$\sigma=\pm 1$. Ferromagnetism is incorporated via a Stoner
model, with an exchange splitting $h_0(\vec l)$ on site $\vec l$.
In the presence of local s-wave pairing described by a
superconducting order parameter $\Delta(\vec l)$, the Bogoliubov
equation takes the form

$$E\psi^\sigma(\vec l) =[\epsilon(\vec l)-\sigma h_0(\vec l)] \psi^\sigma(\vec l) -\sum_{\vec\delta} \gamma
\psi^\sigma(\vec l+\vec\delta)+ \sigma\Delta(\vec l)
\phi^{-\sigma}(\vec l)\eqno{(5)}$$
$$ E\phi^\sigma(\vec l) =- [\epsilon(\vec l)-\sigma h_0(\vec l)] \phi^{\sigma}(\vec l)
+\sum_{\vec\delta} \gamma \phi^\sigma(\vec l+\vec\delta)
-\sigma\Delta^*(\vec l)\psi^{-\sigma}(\vec l),\eqno{(6)}$$ where
$\vec l + \vec\delta$ sums over the nearest neighbors of $\vec l$.

If $\vec l$ belongs to the normal diffusive region of figure 1,
$\epsilon(\vec l)$ is chosen to be a random number, uniformly
distributed over the interval $\epsilon_0 -W/2$ to
$\epsilon_0+W/2$, whereas in the clean F and S regions
$\epsilon_i=\epsilon_0$. In the S region, the order parameter is
set to a constant, $\Delta(\vec l)=\Delta_0$, while in all other
regions, $\Delta(\vec l)=0$. If $\vec l$ is located in one of the
clean ferromagnetic leads, then Stoner splitting is $h_0(\vec l)=
h_j$, where $j=1,2$ labels the lead containing site $\vec l$. For
aligned moments $h_1=h_2=\epsilon_M$ and for anti-aligned moments,
$h_1=-h_2=\epsilon_M$. If $\vec l$ is not located in one of the
ferromagnetic leads then $h_0(\vec l)=0$. The nearest neighbor
hopping element $\gamma$ fixes the energy scale (ie the
band-width), whereas $\epsilon_0$ determines the band-filling. In
what follows, to model the experimentally-relevant regime of
$\Delta_0 \ll {\rm Fermi\,\,\, energy} (=4\gamma -\epsilon_0)$ and
$M_f
> {\rm Fermi\,\,\, wavelength}$, we set $\gamma=1$, $\epsilon_0=0.2$, $\epsilon_M= 3.8$, $M_f=10$
and $\Delta_0= 0.1$.

By numerically solving the Bogoliubov - de Gennes equation for a
given energy $E$, the scattering matrix $S$ can be computed for a
given realization of the disorder and average values of measurable
quantities obtained ensemble averaging the quantity of interest.
For a given value of disorder $W$, the elastic mean free path
$\lambda_{el}$ of the normal region is obtained by from a separate
calculation of the ensemble-averaged dc conductivity of a
rectangular normal bar, connected to normal reservoirs.
$\lambda_{el}$ is then obtained by comparing the result with the
Drude formula. In what follows, the lattice constant is set to
unity and $W$ chosen such that $\lambda_{el}=10$.

For $E=0$, figures 2 and 3 show the dependence on $L$ and $M$ of
the ensemble-averaged Andreev transmission coefficient in the
presence of non-magnetic wires, while figures 4 and 5 show the
corresponding quantities in the presence of fully spin-polarized
wires with opposite magnetizations. For small $L$ (eg $L=1$) and
finite $M$, figures 3 and 5 show that $T_a$ is exponentially
small, as predicted by   \cite{feinberg}. However as $L$ is
increased to a finite value, $T_a$ shows a dramatic enhancement.

Although this structure is difficult to describe analytically,
the qualitative behaviour of $T_a$ is clear and can be understood
in terms of a simple resistor model, in which $T_a$ is
proportional to the conductance of electrons from lead 1
travelling first to the N-S interface and then to lead 2.
Consider for example figure 2, which for $M \ne 0$ shows the
expected exponential suppression of $T_a$ at $L=0$. For small $L$,
$T_a$ increases linearly with $L$. The slope of this linear
regime is independent of $M$ for small $M < \lambda_{el}$,
reflecting the ballistic nature of the N region in this limit,
whereas for the opposite limit of $M
> \lambda_{el}$, the slope decreases with increasing $M$,
reflecting the diffusive behaviour of the N region. In the
diffusive regime, for $M << L$, this leads us to expect the $T_a$
is proportional to $L/(M+2M_f)$, whereas for $L << M$ it is
proportional to $(M+2M_f)/L$, which suggests that the maximum
value of $T_a$ occurs when $L \approx M +2M_f$. Figure 4 shows
that same qualitative behaviour also persists in the presence of
ferromagnetic leads.

In conclusion, we have shown that in contrast with the tunneling
regime, where $T_a$ is exponentially small, the non-local current
can be dramatically enhanced by including a normal region between
the F leads and superconductor. In this case the non-local current
no longer decays monotonically with $M$ or $L$, but instead
exhibits a maximum corresponding to a square normal region with $L
\approx M +2M_f$.

\section*{Acknowledgements} Work supported in part by the European
Community's Human Potential Programme under contract
HPRN-CT-2000-00144, Nanoscale Dynamics.

\begin{figure}[h]
\epsfysize=6cm \epsfxsize=8cm \centerline{\epsffile{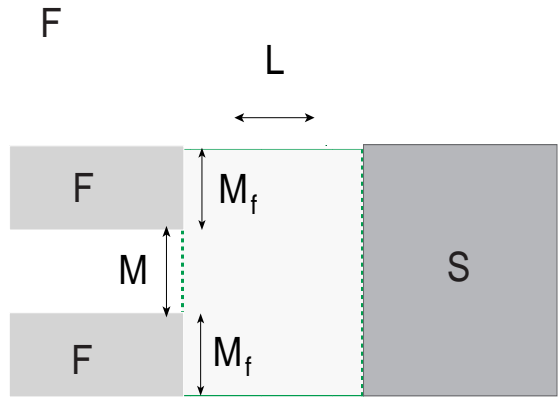}}
\caption{\it \label{geometry-1} The hybrid nanostucture analysed.}
\end{figure}

\begin{figure}[h]
\epsfysize=6cm \epsfxsize=8cm \centerline{\epsffile{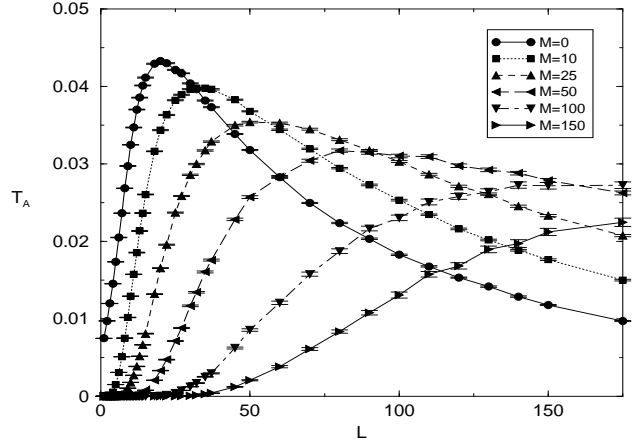}}
\caption{\it \label{geometry-2} For normal leads (ie when
$\epsilon_M=0$) this figure shows Andreev transmission against
length $L$ for different contact separations $M$.}
\end{figure}

\begin{figure}[h]
\epsfysize=6cm \epsfxsize=8cm \centerline{\epsffile{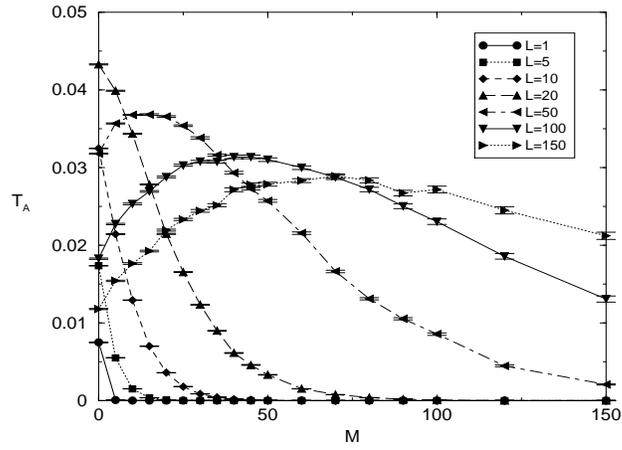}}
\caption{\it \label{geometry-3} For normal leads, this figure shows
Andreev transmission against contact separation $M$ for different
lengths $L$.}
\end{figure}
\begin{figure}[h]
\epsfysize=6cm \epsfxsize=8cm \centerline{\epsffile{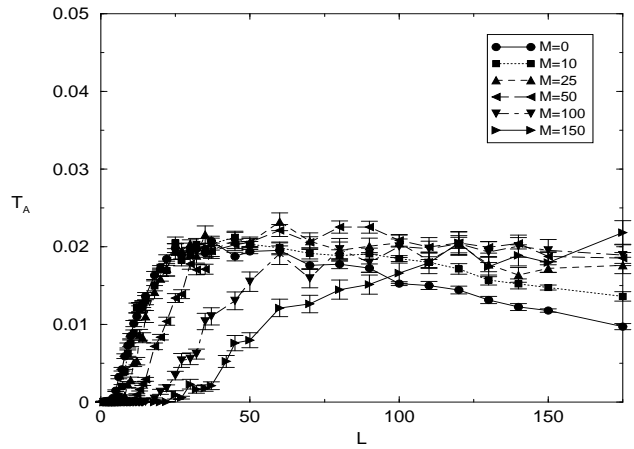}}
\caption{\it \label{geometry-4} For ferromagnetic leads (ie
$\epsilon_M=3.8$) with anit-aligned magnetiszations, this figure
shows Andreev transmission length $L$ for different contact
separations $M$.}
\end{figure}
\begin{figure}[h]
\epsfysize=6cm \epsfxsize=8cm \centerline{\epsffile{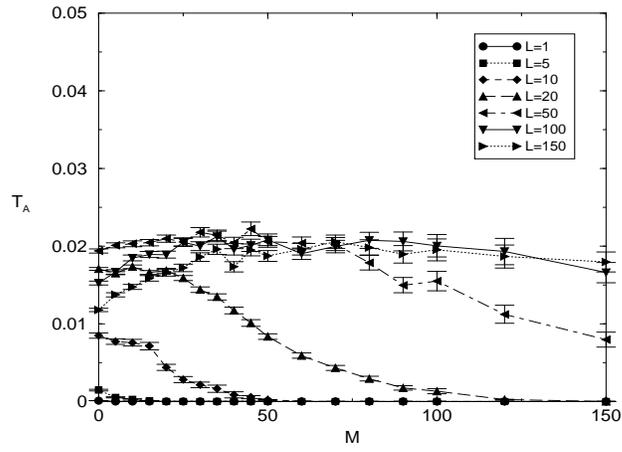}}
\caption{\it \label{geometry} For ferromagnetic leads (ie
$\epsilon_M=3.8$) with anit-aligned magnetizations, this figure
shows Andreev transmission against contact separation $M$ for
different lengths $L$.}
\end{figure}



\end{document}